# Hydrodynamic Insight Drives Multimodal Light-Field Dynamics via Streamline Engineering

*Wenxiang Yan,[1,2,†] Zheng Yuan,[1,2,†] Yuan Gao,[1,2] Zhaozhong Chen,[4] Zhi-Cheng Ren,[1,2] Xi-Lin Wang,[1,2] Jianping Ding,[1,2,3,*] and Hui-Tian Wang[1,2,*]*

[1]National Laboratory of Solid State Microstructures and School of Physics, Nanjing University, Nanjing 210093, China

[2]Collaborative Innovation Center of Advanced Microstructures, Nanjing University, Nanjing 210093, China

[3]Collaborative Innovation Center of Solid-State Lighting and Energy-Saving Electronics, Nanjing University, Nanjing 210093, China

[4]James Watt School of Engineering, University of Glasgow, Glasgow, G12 8QQ, UK

[†]These authors contributed equally to this work.

*Corresponding author: jpding@nju.edu.cn; htwang@nju.edu.cn

## Abstract

Since the 1970s, analogies between laser dynamics and fluid systems have provided insight into phenomena such as chaos, multistability, and turbulence. Building on this perspective, we model the optical field as an energy "fluid" and interpret Poynting-vector trajectories as "energy streamlines," yielding a unified, three-dimensional map of light's free-space dynamics. By sculpting these streamlines, we develop an approach to talior vortex-beam propagation dynamics that suppresses both diffraction- and OAM–induced broadening. Extending this method to general structured modes, we enable a single field to exhibit customizable multimodal dynamics that integrate features from primary structured light families: the diffraction-free, self-healing behavior of Bessel beams; the tunable self-similarity of Laguerre–Gaussian beams and adjustable self-acceleration of Airy beams. Additionally, it allows for adjustable propagating energy-density profiles to counteract losses. Optical-tweezer experiments—analogous to particle-tracking velocimetry in fluid dynamics—show that trapped microspheres closely follow the designed streamlines, validating the streamline geometries and indicating a potential route toward precision 3D optomechanical control. In a proof-of-principle free-space communication experiment, vortex beams with customized multimodal dynamics demonstrate several improvements, including more independent channels, reduced turbulence-induced mode scattering, and robust non-line-of-sight transmission. Together, the streamline-engineering approach offers a unified and adaptable strategy for tailoring light's propagation dynamics, with potential applications in precision optomechanics, optofluidics, and advanced optical networking.

## Introduction

Structured light beams[1,2], as exact solutions of the Helmholtz equation, exhibit a rich variety of free-space dynamics. Gaussian-mode beams[3,4]—including fundamental, Hermite–Gaussian, Laguerre–Gaussian and Ince–Gaussian modes—display self-similar diffraction, preserving their transverse profiles up to overall scaling. By contrast, Bessel beams[5] are diffraction-free and self-healing, inspiring cosine[6], Mathieu[7] and parabolic[8] variants that reconstruct after perturbation. Airy beams, endowed with a caustic phase[9], unite non-diffraction and self-healing with transverse parabolic acceleration[10,11], spawning a spectrum of paraxial and non-paraxial accelerating[12] and autofocusing modes[13]. Despite these advances, classical modes obey fixed propagation laws (for example, Airy beams accelerate along a predetermined parabola and Gaussian modes diverge hyperbolically), which limits their adaptability. While specialized techniques[14-17] can tune individual propagation behaviors, they typically address only one behavior at a time and are incompatible with each other. *Consequently, developing theoretical frameworks and adaptive strategies that enable the simultaneous customization of multiple propagation characteristics is essential to unlock the full potential of structured light[1,2] in complex, real-world applications.* To illustrate how these propagation limitations manifest in practice—and why it matters for real-world uses—we now turn to vortex beams as a concrete case study.

Among structured light forms, vortex beams[18,19]—exemplified by higher-order Laguerre–Gaussian modes— represent a distinct subclass whose unique propagation dynamics present further challenges. Such beams carry orbital angular momentum (OAM) with a helical wavefront and central dark core, and have attracted growing interest in optical manipulation, communications, and quantum technologies[20,21]. However, their practical deployment remains hindered by two coupled propagating defects: *First, beams with higher OAM orders exhibit substantial radial expansion and divergence, posing challenges for spatial-division multiplexing, fiber transmission, and optical trapping*[22-24]. *Second, conventional vortex fields are subject to diffraction-induced broadening over extended propagation distances, degrading their performance in long-range communication, sensing, imaging, and quantum information*[25-27]. For instance, the beam radius of a zero–radial-index Laguerre–Gaussian mode increases with both topological charge ($l$) and distance ($z$), following: $r(|l|, z) = w_0\sqrt{|l|+1}\sqrt{1+(z/z_0)^2}$, where $w_0$ and $z_0$ denote the beam waist and Rayleigh distance. Although specialized designs — including perfect vortex[28], Bessel vortex[5], multi-vortex geometric[29] and iso-propagation beams[30,31] — can mitigate one of these issues, *none concurrently overcomes both.*

In this work, we introduce a hydrodynamic framework for multimodal control of vortex-beam propagation dynamics. By exploiting the analogy between fluid flows and electromagnetic fields, we adopt energy streamlines—3D trajectories of the Poynting vector—as intuitive descriptors of photon motion and energy transport in free-space propagation (Section 1). Building on this insight, we reverse the perspective: through deliberate engineering of these streamlines, we develop an approach to talior vortex-beam propagation dynamics that simultaneously suppresses both diffraction- and OAM–induced broadening, yielding a true non-diffracting perfect vortex whose ring diameter remains invariant with topological charge $\ell$ and propagation distance $z$ (Section 2). We then extend this deterministic propagation control from vortex beams to general structured modes, endowing them with customizable multimodal dynamics that combine hallmark features of classical structured light: the diffraction-free, self-healing behavior of Bessel beams; OAM-invariant evolution of iso-propagation modes; the tunable self-similarity of Laguerre–Gaussian beams and adjustable self-acceleration of Airy beams, in contrast to the fixed properties of their classical counterparts; and controlled energy delivery to mitigate attenuation (Section 3). Optical-tweezer experiments—analogous to particle-tracking velocimetry in fluid dynamics—confirm that trapped microspheres faithfully follow the designed streamlines, validating the hydrodynamic model and suggesting a route to precision 3D optomechanical control (Section 4). In a proof-of-principle free-space communication experiment, vortex beams with customized multimodal dynamics demonstrate a significant increase in independent channels, enhanced turbulence resilience, and improved non-line-of-sight capability (Section 5). Overall, this hydrodynamic streamline-engineering toolkit provides an experimentally validated, versatile platform for tailoring multimodal propagation of structured light—opening new opportunities in precision optomechanics, advanced optofluidics, and next-generation optical communications.

# Results

## 1. Mapping Light Propagation through Streamline Dynamics

**Fluid–optics analogy.** The parallel between laser dynamics and fluid or superfluid behavior dates back to the early 1970s, when laser-physics equations were recast into the form of the complex Ginzburg–Landau equations[32]. Since then, this hydrodynamic framework has provided deep insight into a host of phenomena—from superconductivity and superfluidity to Bose–Einstein condensation[33]. Motivated by these connections, researchers have probed the hydrodynamic character of optical fields, revealing a wealth of nonlinear behaviors—chaos, multistability, and even turbulence—predicted theoretically and confirmed experimentally in laser systems[34–39]. In 1989, Coullet et al., inspired by fluid vortices, formally introduced the notion of optical vortices[19], catalyzing the modern exploration of structured light. Building on this hydrodynamic insight, we now broaden the fluid–optics analogy to describe light propagation through the lens of flow fields and streamlines.

**Streamline in fluid dynamics.** In classical fluid dynamics, the Eulerian description characterizes the fluid's motion via a velocity field $v(R, t)$, where $R = \{x, y, z\}$ denotes spatial position. By integrating this field with the hydrodynamic differential equations, one visualizes the flow using streamlines—trajectories whose tangent at each point aligns with the local velocity. Formally, if $R(z, t) = \{x(z, t), y(z, t), z\}$, then its evolution obeys

$$\frac{dx(z,t)}{v_x(R(z),t)} = \frac{dy(z,t)}{v_y(R(z),t)} = \frac{dz}{v_z(R(z),t)} \tag{1}$$

where $v = \{v_x, v_y, v_z\}$. Regions where streamlines cluster correspond to faster flow, whereas sparse regions signify slower motion. This formalism lays the groundwork for mapping optical-field evolution onto an analogous hydrodynamic picture.

**Streamline for Light Propagation.** In this analogy, a monochromatic light field can be treated as an steady-state unchanging flow field[40] ($\partial v/\partial t = 0$), *where its "velocity field" is represented by the momentum distribution of light*—quantified by the Poynting vector[41]. Specifically, for a scalar wave $\psi(R)$, the Poynting vector $p(R)$—the local expectation value of the momentum operator—is given by

$$p(R) = \mathrm{Im}\, \psi^*(R) \nabla \psi(R) = |\psi(R)|^2 \nabla \arg \psi(R), \tag{2}$$

By substituting this momentum distribution (Poynting vectors) into Eq. (1), the energy streamlines of the optical field — "roadmap" of photon motion —can be derived, which is the streamlines of the Poynting vector. These trajectories, $R(z) = \{x(z), y(z), z\} = \{r(z), \varphi(z), z\}$, can be determined in Cartesian or cylindrical coordinates by solving the hydrodynamic differential equations[40]:

$$dx(z)/dz = p_x(R(z))/p_z(R(z)), \quad dy(z)/dz = p_y(R(z))/p_z(R(z)); \tag{3-1}$$

$$dr(z)/dz = p_r(R(z))/p_z(R(z)), \quad d\varphi(z)/dz = p_\varphi(R(z))/[r(z)p_z(R(z))]; \tag{3-2}$$

where $\boldsymbol{p} = \{p_x, p_y, p_z\} = \{p_r, p_\varphi, p_z\}$. These energy streamlines offer an intuitive picture of light propagation, often likened to the "Bohmian trajectories" that describe experimentally measurable paths of average photon trajectories[40,42–44]. In quantum physics, the trajectories of the Poynting vector in light (or quantum-mechanical waves) are described as streamlines in the Madelung hydrodynamic interpretation[45], which are later regarded as paths of quantum particles in the Bohm–de Broglie interpretation[46,47].

For Bessel vortex beams[40], the energy streamlines follow the helical trajectories

$$\boldsymbol{R}(z) = \{r(z), \varphi(z), z\} = \left\{ r_0, \varphi_0 + \frac{l}{r_0^2 k_{z0}} z, z \right\}, \tag{4}$$

where $r_0$ and $\varphi_0$ denote each streamline's initial radial and azimuthal coordinates, $k_{z0}$ represents the longitudinal wavevector component of the plane waves that constitute the Bessel beam. *These paths wrap indefinitely around a cylinder of initial radius $r_0$ with no radial divergence (Fig. 1C).* In contrast, Laguerre–Gaussian beams yield streamlines of the form[40]

$$\boldsymbol{R}(\zeta) = \{\rho(\zeta), \varphi(\zeta), \zeta\} = \left\{ \rho_0 \sqrt{1+\zeta^2}, \varphi_0 + \frac{l}{\rho_0^2} \arctan \zeta, \zeta \right\}, \tag{5}$$

in scaled coordinates $(r, \varphi, z) \equiv (\omega_0 \rho, \varphi, k\omega_0^2 \zeta)$, where $k$ donates the wavenumber. *Here, each trajectory traces a helix on a hyperboloid, exhibiting pronounced radial spreading (Fig. 1A).* These helical energy streamlines vividly illustrate the intrinsic rotational flow characteristic of vortex beams, serving as the hydrodynamic manifestation[48] of optical OAM.

Reversing the perspective in Eqs. (2–5), we introduce a four-step streamline-engineering approach (Methods 1-3) to tailor the propagation dynamics of light:

1. **Streamline Configuration**: Prescribe the desired trajectories $\boldsymbol{R}(z)$ to encode specific propagation behaviors.
2. Momentum-field Sculpting**.** Use the hydrodynamic equations [Eq. (3)] together with the fluid-continuity condition (Method 3) to compute the required momentum distribution $\boldsymbol{p}(\boldsymbol{R})$.
3. **Angular-spectrum engineering.** Devise an angular-spectrum distribution $S(k_r, \phi, k_z)$ in momentum-space that produces the target momentum field when transformed into real space.
4. **Real-space beam construction:** Implement an optical Fourier transform (via a lens) to convert $S(k_r, \phi, k_z)$ into the physical field $\psi(\boldsymbol{R})$, realizing deterministic control over its multimodal dynamics.

Building on this streamline-engineering approach, we achieve deterministic control of vortex-beam propagation dynamics, simultaneously suppressing diffraction- and OAM-induced broadening (Section 2). We then generalize this strategy to arbitrary structured modes, equipping them with customizable multimodal dynamics that combine the hallmark features of classical structured light (Section 3).

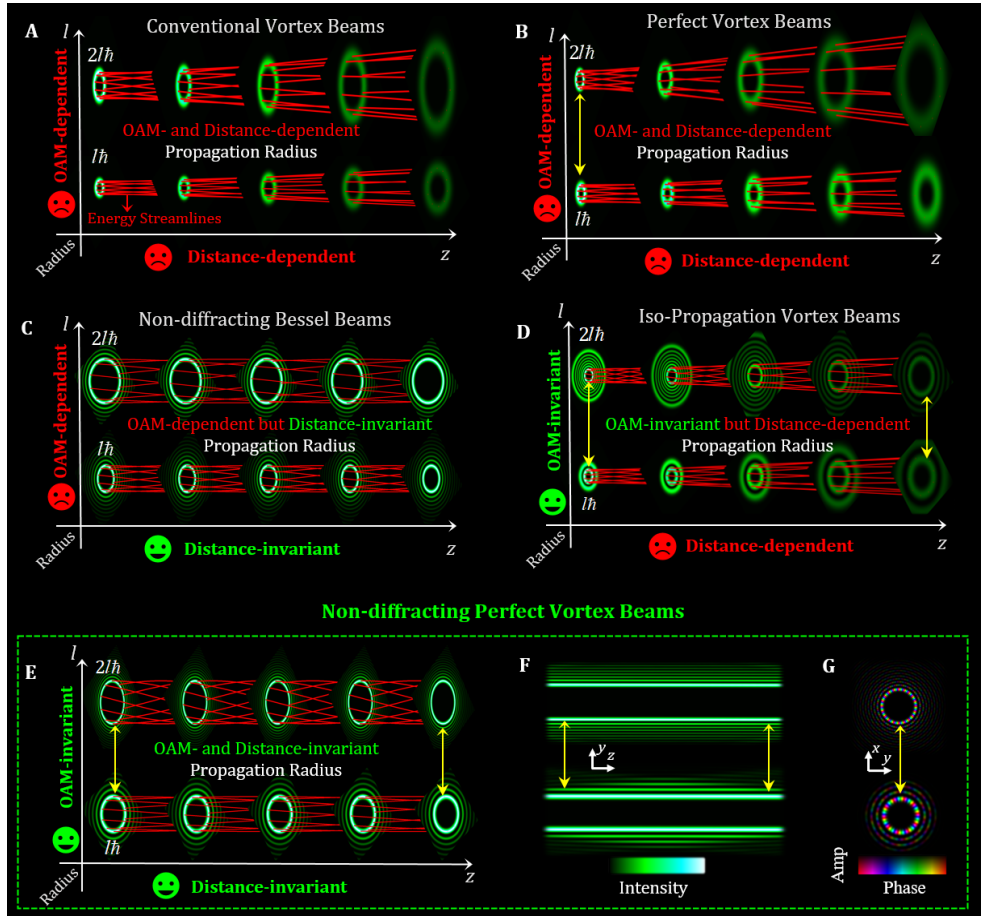

**Fig. 1. Hydrodynamic Depiction for Vortex-beam Propagation Dynamics.** (**A**) Conventional vortex beams and (**B**) Perfect vortex beams—radii increase with both OAM and distance; rotation speed of energy streamlines (red curves) slow down over propagation and do not speed up with higher OAM. The yellow double-headed arrow in (**B**) indicates that perfect vortex fields merely maintain OAM-invariant radii in a single transverse plane. Here, $l\hbar$ ($2l\hbar$) denotes the OAM per photon. (**C**) Non-diffracting Bessel beams—radius grows with OAM but stays constant over distance; rotation speed of energy streamlines decelerates as OAM increases but is invariant along propagation. (**D**) Iso-propagation vortex beams—radius is fixed for all OAM (yellow arrows) but expands with distance; rotation speed of energy streamlines accelerates synchronously with OAM yet decreases as the beam propagates. (**E**) Non-diffracting perfect vortex beams (NDPVBs)—radius is invariant of both OAM and distance (yellow arrows); rotation speed of energy streamlines accelerates synchronously with OAM and remains constant during propagation. (**F**) y-z intensity maps and (**G**) x-y complex-amplitude distributions in plane of the NDPVBs from (**E**) with $l$ = 10. The streamlines herein are directly drew from the "streamline" function with the distributions of Poynting vector caculated in Matlab, which are consistent with the analytical solutions from the hydrodynamic differential equations in this work. The demonstration for the streamline rotation speeds can be found in Supplementary Note 4.

## 2. Streamline Engineering: Overcoming Dual Propagation Limits of Vortex Beam

**Propagating behaviors of Existing Vortex Beams.** Conventional vortex beams face two coupled propagation challenges: OAM-induced expansion and diffraction-driven spreading (Fig. 1A: OAM- and distance-dependent propagation radius). These dual issues significantly limit the performance of vortex beams in applications such as optical multiplexing, fiber-optic communication, particle manipulation, and long-distance quantum communication[22,23,27]. To date, researchers have developed specialized beam designs that each mitigate one of these problems. For example:

- Perfect vortex beams[28,31,49] maintain a OAM-independent ring size in a single transverse plane, but their beam radius still changes with propagation distance (Fig. 1B).
- Non-diffracting Bessel beams[5] have a propagation-invariant intensity profile (due to their zero divergence), yet their ring radius grows substantially with increasing OAM (Fig. 1C).
- Iso-propagation vortex beams[30,31] are engineered to keep a constant radius regardless of OAM, but they still undergo diffraction-induced broadening as they propagate (Fig. 1D).

As existing vortex beam designs have yet to simultaneously overcome both diffraction-induced spreading and OAM-driven expansion, we turn to a hydrodynamic perspective to gain deeper insight into their coupled propagation dynamics.

**Hydrodynamic Perspective of Energy Rotaion.** In this hydrodynamic view, the beam's energy flows along helical streamlines with an angular rotation rate $\omega_z = d\varphi(z)/dz = \varphi'(z)$. This rotation rate reflects the photon's OAM[48] and also influences optical forces and momentum exchange during light–matter interactions[40,50]. In practice, the combined effects of diffraction-driven spreading and OAM-driven expansion cause $\omega_z$ to diminish as either the propagation distance or the OAM increases (see energy streamlines in Fig. 1 and details demonstrations Supplementary Note 4). In other words, because the beam's radius grows with both distance and OAM, the energy rotation rate does not increase for higher OAM; instead, it slows down over propagation. These observations motivate a new strategy: by synchronizing $\omega_z$ with the beam's OAM and stabilizing it against diffraction-induced decay, beam broadening might be simultaneously suppressed. Building on this insight, we now engineer the streamline geometry to configure a novel vortex beam whose energy rotation rate remains constant during propagation and scales with OAM, yielding ideal free-space vortex dynamics.

**Non-diffracting Perfect Vortex Beam (NDPVB).** To eliminate both diffraction- and OAM-induced broadening at once, we sculpt the energy streamline geometry from non-diffracting Bessel beams (Eq. 4) to the new form $\boldsymbol{R}(z) = \{r_I, \varphi_0 + l/(r_I^2 k_{zl})z, z\}$, yielding a non-diffracting perfect vortex beam (NDPVB), where $r_I$ is the fixed radius of the beam's innermost ring and $k_{zl}$ is a custom longitudinal wavevector component. In this construction, the subscript "I" signifies that $r_I$ is invariant: the NDPVB's transverse ring size remains constant for any topological charge $l$ and any propagation distance $z$ (see Figs. 1E–G). Following the four-step streamline-engineering approach (Methods 1-2), the reconstructed angular-spectrum distribution of a NDPVB with a non-diffracting range $z \in (b-a, b+a)$ can be expressed as

$$S(k_x, k_y) = \mathcal{F}_z\{rect[(z-b)/2a]e^{ik_{zl}z}e^{il\varphi}\} = sinc(2a(k_z - k_{zl}))e^{-ik_z b}, \qquad (6)$$

where the subscript "z" of "$\mathcal{F}$" refers to the Fourier transform dimensions and $k_x^2+k_y^2+k_z^2=k^2$. The role of $k_{zl}$ is to precisely offset the usual OAM-driven expansion of Bessel-like modes, anchoring the beam's propagation-invariant radius at $r_l$ regardless of $l$ (Methods 1-2 with Supplementary Notes 1). As a result of this design, the NDPVB's energy rotation rate becomes $\omega_z = l/(r_l^2 k_{zl}) \approx l/(r_l^2 k) \propto l$, which scales linearly with the OAM $l$ and remains constant during propagation. In other words, $\omega_z$ synchronizes with the beam's OAM and does not decay over distance, exhibiting ideal free-space vortex dynamics (illustrated in Fig. 1E). Unlike conventional "perfect" vortex beams (which only produce an OAM-independent ring size in one transverse plane and degrade rapidly off that plane), these Non-Diffracting Perfect Vortex Beams (NDPVBs) are the first truly three-dimensional perfect vortices with OAM- and propagation-invariant size (Figs. 1E–G), simultaneously overcoming the dual propagation limits of diffraction and OAM-induced expansion. (Experiment validations in Supplementary Note 1)

## 3. Multimodal Dynamics Tailoring by Streamline Sculpting

Building on our success in overcoming the dual propagation limits of vortex beams (Section 2), we now extend streamline engineering to the broader realm of structured light. In the Introduction, two tiers of challenge have emerged—first, the need for multidimensional propagation-dynamics control across diverse beam types, and second, the specific shortcomings of simple vortex dynamics. Structured light modes such as Airy beams (which enable precise self-accelerated steering), Gaussian beams (with tunable spot size for adaptive focusing), and Bessel beams (offering diffraction-free, self-healing transmission) each exhibit distinctive, application-enabling propagation behaviors[1,51]. Yet vortex beams historically lack this multifunctionality. By adopting our hydrodynamic framework and sculpting energy-streamline trajectories, we can now imprint these varied dynamics—the diffraction-free, self-healing behavior of Bessel beams; the tunable self-similarity of Laguerre–Gaussian variants and adjustable self-acceleration of Airy variants, in contrast to the fixed properties of their classical modes; and controlled energy delivery to mitigate environment attenuation—onto Non-Diffracting Perfect Vortex Beams (NDPVBs). These results suggest that streamline sculpting may offer a flexible strategy to extend vortex-beam functionality and informing broader structured-light control.

**Sculpting Streamlines for Adjustable Self-Similar Propagation.** In cylindrical coordinates, the NDPVB's streamlines are reshaped from $\boldsymbol{R}(z) = \{r_l, \varphi_0 + l/(r_l^2 k_{zl})z, z\}$ to $\{r_l(z), \varphi_0 + l/[r_l^2(z)k_{zl}(z)]z, z\}$, thereby endowing the beam with a dynamically adjustable self-similar radius $r_l(z)$ (Figs. 2A–B). The resulting configuration permits precise regulation of the rotational rate during propagation, which is given by $\omega_z(z) = l/[r_l^2(z)k_{zl}(z)] \approx l/[r_l^2(z)k]$.

**Sculpting Streamlines for Customized Self-Accelerating Trajectories.** In Cartesian coordinates, the original streamlines $\boldsymbol{R}(z) = \{r_l \cos\varphi(z), r_l \sin\varphi(z), z\}$ are translated to $\{x_s(z) + r_l \cos\varphi(z), y_s(z) + r_l \sin\varphi(z), z\}$, where $\varphi(z) = \varphi_0 + l/(r_l^2 k_{zl})z$, so that propagation follows an arbitrary three-dimensional trajectory $(x_s(z), y_s(z))$, thus imparting customizable self-accelerating dynamics to the NDPVB (Figs. 2C–D).

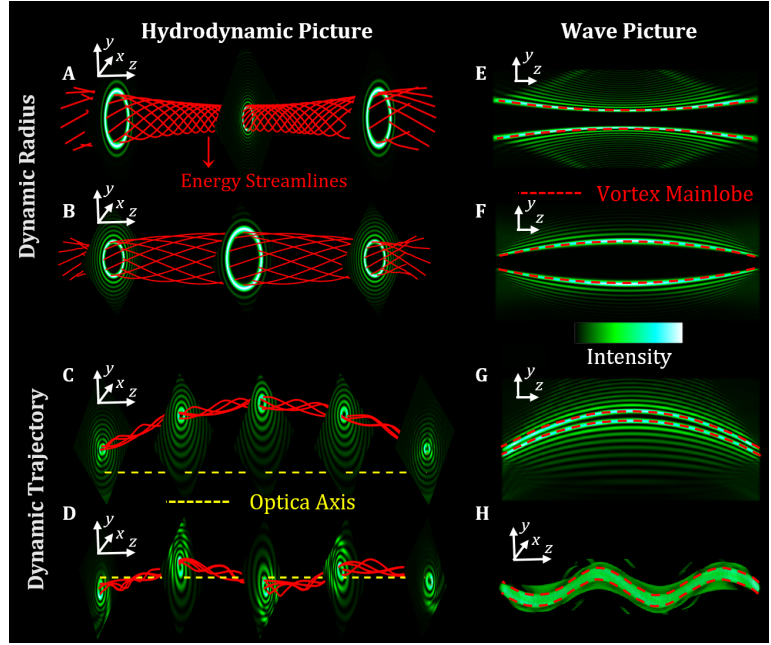

**Fig. 2. Sculpting Streamlines for adjustable self-similarity and customized self-acceleration. (A, B)** NDPVBs with adjustable self-similarity, exhibiting self-shrinking (**A**) and self-stretching (**B**) radii; red curves indicate the energy streamlines along the vortex mainlobe. (**C, D**) NDPVBs with customizable self-accelerating dynamics, propagating along a parabolic (**C**) and a spiral (**D**) trajectory. (**E–G**) Intensity maps of (**A–C**) in the y–z plane; red dashed lines denote the vortex mainlobes. (**H**) Three-dimensional intensity iso-surface corresponding to (**D**).

**Streamline Interactions Enabling Self-Healing Propagation.** NDPVBs, whose innermost-ring radius remains invariant with respect to topological charge $l$ and propagation distance $z$, are decomposed into two regions: the mainlobe (i.e., region of interest), which carries the highest intensity and governs interactions with matter or detectors, and a series of surrounding sidelobes. The sidelobes sustain and regulate the propagation dynamics of the mainlobe; together, they constitute the entire beam and account for its total energy. During free-space propagation, main- and sidelobe energy streamlines rotate along their respective cylindrical surfaces without mutual energy exchange. When an obstacle intercepts and truncates the main-lobe streamlines, the sidelobe streamlines spontaneously flow into the mainlobe region upon further propagation, reconstructing both its intensity profile and streamline structure (Fig. 3A). In the absence of these supportive sidelobe streamlines in conventional vortex beams, such spontaneous self-healing cannot occur (see Supplementary Note 5 and Supplementary Movie 1 for details).

**Manipulating Streamline Interactions for Tunable Energy Transfer.** During self-healing, sidelobe streamlines converge into the mainlobe without tunability. By invoking the optical "fluid continuity equation" (Method 3), these sidelobes can be recast as an energy reservoir for the vortex mainlobe. The rate of sidelobe-to-mainlobe streamline transfer can be actively controlled, thereby dynamically adjusting the mainlobe's spatial energy density to satisfy application-specific intensity requirements. For instance, by directing nearly all sidelobe streamlines into the mainlobe, the local energy density associated with matter or detector interactions can be selectively enhanced (Fig. 3C), yielding improved energy efficiency and signal-to-noise ratios (a case study in Supplementary Note 6 demonstrates a threefold improvement, as well as corresponding experimental results of Supplementary Movie 2). Moreover, by tuning the sidelobe inflow rate, the mainlobe intensity can be held constant across media with different attenuation coefficients, thereby meeting the stringent requirements of imaging and sensing in complex environments under uniform structured-light illumination[52]. In Figs. 3G–J, attenuation profiles were first measured using a fundamental NDPVB in various decay media, and the corresponding optimal sidelobe convergence rates were determined to dynamically compensate for the main-lobe's energy decay. (Experiment Setup in Supplementary Note 7)

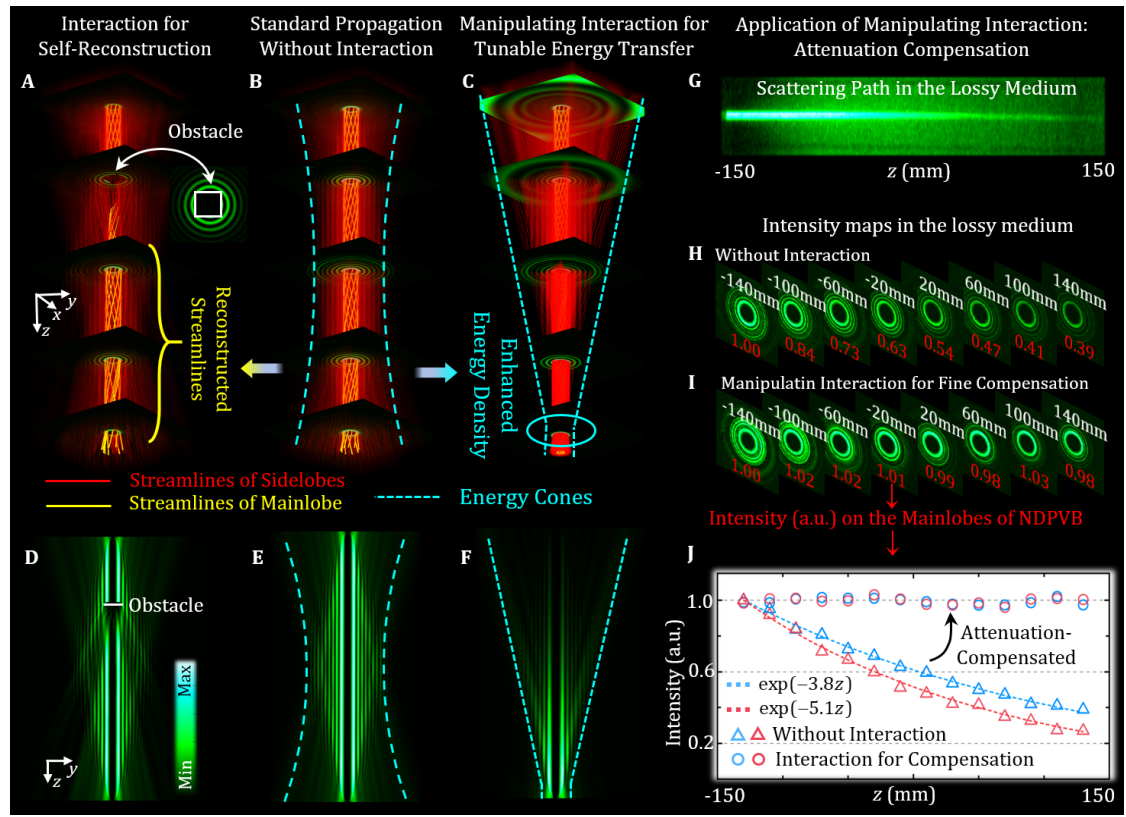

**Fig. 3. Manipulating Streamline Interactions for Self-Healing and Tunable Energy Delivery. (A)** When the mainlobe of a non-diffracting perfect vortex beam (NDPVB) is intercepted by an obstacle (white rectangle), its yellow energy streamlines are truncated. The sidelobe streamlines (red curves) spontaneously flow into the mainlobe region, reconstructing the original streamline structure. **(B)** In free-space propagation, a standard NDPVB exhibits no interaction between mainlobe and sidelobe streamlines. **(C)** By actively converging sidelobe streamlines into the mainlobe, the local spatial energy density is enhanced; blue-dotted curves highlight the primary region of energy concentration. **(D–F)** Intensity maps in the y–z plane corresponding to (**A–C**), respectively. **(G)** Attenuated scattered path of a standard NDPVB propagating in a lossy medium (milk suspension). **(H)** Propagating profiles of the standard NDPVB in the lossy medium, with red font digits indicating the normalized intensity on the mainlobe (blue triangles in (**J**)), used to probe the attenuation curve(blue-dashed curve in (**J**)). **(I)** After adjusting the inflow rate of sidelobe streamlines, the main-lobe intensity remains invariant (blue circles in (**J**)) despite the medium's attenuation. For a second lossy medium (red-dashed curve in (**J**)), the probing and dynamically compensated mainlobe intensities are shown by red triangles and red circles,

respectively. **(J)** Attenuation curves for both media (blue- and red-dashed) with corresponding normalized main-lobe intensities: triangles denote unmodified probing NDPVBs and circles denote dynamically compensated NDPVBs.

In contrast to classical modes, which follow fixed propagation laws and often require specialized techniques[14–17] that address single propagation behaviors in isolation, the streamline-engineering approach offers a more flexible and unified strategy for controlling the multimodal propagation dynamics of structured light. This hydrodynamic framework not only addresses the dual challenges of diffraction- and OAM-induced broadening in vortex beams, but also provides the ability to endow single beam with customizable multimodal propagation dynamics that combine hallmark features of classical structured light: the diffraction-free, self-healing behavior of Bessel beams; OAM-invariant evolution of iso-propagation modes; the tunable self-similarity of Laguerre–Gaussian beams; and adjustable self-acceleration of Airy beams. Additionally, it enables controlled energy delivery to compensate for media attenuation. These findings suggest that the streamline-engineering method could serve as a versatile toolkit for tailoring propagation dynamics across different beam modalities, potentially broadening the functional applications of structured light[1,2,51,53]. The following sections will explore the potential benefits of these streamline-engineered vortex beams, particularly in optical manipulation (Section 4) and free-space optical communications (Section 5).

## 4. Optical Tweezers as Particle-Tracking Velocimetry: Probing Streamline Dynamics

In classical particle-tracking velocimetry for fluid dynamics[54], tracer beads are injected into a flow and their trajectories are recorded by high-speed cameras to reconstruct the underlying streamlines. An analogous strategy was adopted here: microspheres were seeded into the shaped light field, and their motion was captured to map the beam's "streamlines". The optical-tweezers[55] platform (Fig. 4A) was built on an inverted confocal microscope (Nikon TE2000-U) equipped with a 4-f holographic beam-shaping stage housing a reflective spatial light modulator (Holoeye Leto). A continuous-wave laser (Coherent Verdi-V5) was phase-modulated with computer-generated holograms to generate the required multimode vortex fields. These fields interacted with polystyrene microspheres of 2 μm diameter suspended in de-ionized water inside a custom open-top chamber, which comprised an acrylic plate with a through-hole and a glass coverslip forming the base. *The open geometry minimised hydrodynamic resistance and allowed unrestricted particle motion at the water surface. The chamber was mounted on a Nano-LP200 piezo nanopositioner that displaced the liquid-borne microspheres along the optical axis in synchrony with the camera exposure. This coordinated z-scan permitted three-dimensional bead trajectories to be reconstructed with high spatial accuracy.*

Representative measurements (Figs. 4B–D) reveal that tracer-sphere trajectories in 22nd- and 38th-order NDPVBs, as well as in self-similar NDPVB with the linearly expanding radius, adhere closely to the predesigned energy-streamline contours. By contrast, conventional perfect-vortex beams (Fig. 1B) yield a 2D "perfect" ring only within a single transverse plane; once the beam diffracts off that plane, trapped beads escape and trapping ceases (Fig. 4E). In the NDPVB configuration, however, robust three-dimensional trapping and steering within OAM-invariant radii persist over the full axial scan (Figs. 4B and 4C), thereby validating the propagation-invariant, three-dimensional perfect vortex configuration. Looking ahead, the combination of smaller probe particles, faster volumetric imaging, and axial-view detection modules[56] could achieve improved resolution of energy streamlines. These improvements may enable three-dimensional optomechanical control[51,55] via engineered photon trajectories, potentially extending to applications such as microfluidic flow steering, directed cell transport, and assembly of microscale structures[57].

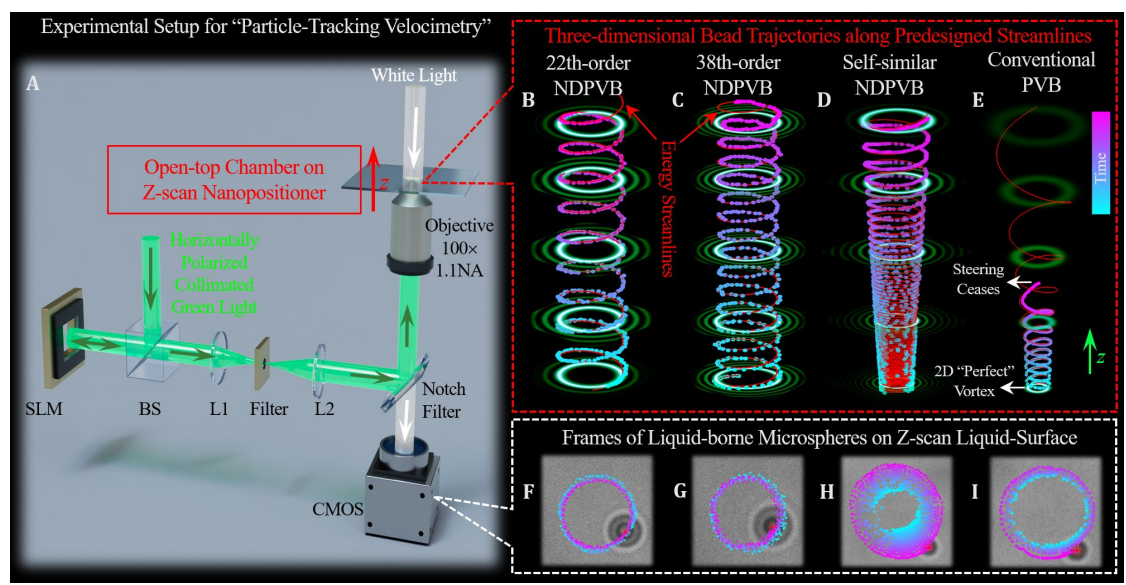

**Fig. 4. Probing Photonic Streamline Dynamics through Optofluidic Velocimetry.** (**A**) Schematic of the optical tweezers platform for photonic streamline mapping. Key components: beam splitters (BS), phase-only spatial light modulator (SLM), and lens pair (L1-L2). The open-top microfluidic chamber, containing 2 μm polystyrene microspheres in deionized water, was mounted on a Nano-LP200 piezo nanopositioner. Synchronized z-axis displacement and CMOS camera exposure enabled 3D trajectory reconstruction of liquid-borne probing microspheres through coordinated scanning. (**B-E**) Reconstructed 3D microsphere trajectories (time-color-coded discrete points) in: (**B**) 22th-order NDPVB, (**C**) 38th-order NDPVB, (**D**) self-similar NDPVB with linear radial expansion, and (**E**) conventional perfect vortex beam. Red curves: Predesigned energy streamlines. The conventional beam (**E**) only maintains 2D "perfect" in a single transverse plane, particles escape and steering ceases when diffracting beyond this plane. (**F-I**) the CMOS camera frames (Supplementary Movie 3) of liquid-borne probing microspheres on z-scan liquid-surface.

## 5. Streamline-Engineered Vortex Beams: Boosting Capacity and Robustness in Free-Space Communications

In free-space optical communication, NDPVBs—whose transverse size does not vary with OAM order or propagation distance—can support a substantially larger number of mutually orthogonal sub-channels than conventional OAM modes (zero-radial-index Laguerre–Gaussian beams), potentially increasing information capacity (Figs. 5A–5C). When considering the effects of atmospheric turbulence on long-haul free-space links and remote sensing and ranging scenarios, NDPVBs were found to experience exhibit relatively weaker and more uniform modal scattering than conventional vortex beams—an observation attributed to their constant, OAM-invariant beam profile over distance (Figs. 5D–5H). By exploiting the customizable self-similar and self-accelerating dynamics of the multimodal NDPVB family, a traditional line-of-sight free-space link can be extended to support robust non-line-of-sight scenarios (Fig. 6). Together, these features suggest a potential framework for next-generation OAM-based free-space optical systems[22,23] that combine higher capacity with turbulence resilience and flexible link geometries.

**Enhance Channel Capacity.** Recent advances in information acquisition and processing have underscored the critical role of multiplexing in scaling communication capacity[58]. Optical multiplexing strategies exploiting polarization and wavelength degrees of freedom have significantly extended system bandwidth[59,60]. Among emerging approaches, spatial mode-division multiplexing—where orthogonal spatial modes act as independent channels—has garnered increasing attention[29,61,62]. In a representative scheme, a free-space optical link combining $Q$ orthogonal OAM modes, two polarization states, and $T$ wavelengths yields an aggregate capacity of $Q\times2\times T\times100$ Gbit/s when each channel carries 100 Gbit/s using quadrature phase-shift keying. This configuration enables petabit-per-second-scale throughput[62], thereby substantially improving both capacity and spectral efficiency in free-space optical communication systems. Despite its promise, vortex-based spatial mode-division multiplexing encounters notable limitations during free-space propagation. Conventional vortex beams suffer from diffraction-induced spreading and OAM-dependent radial expansion—effects that intensify with increasing topological charge and propagation distance[22,23,25,26,63]. As a result, the practical number of accessible spatial subchannels, $Q$, is restricted by both the finite aperture of the receiving optics and these propagation limitations[22,23] (Fig. 5A).

To enable a direct comparison among diverse spatial multiplexing schemes, the number of supported individual subchannels, $Q$, was evaluated as a function of the system-quality factor[64] $S = \pi R_0 \times \text{NA}/\lambda$, where $R_0$ and NA are the common aperture radius and numerical aperture of both transmitter and receiver, and $\lambda$ is the wavelength. This dimensionless factor represents the maximum space–bandwidth product a beam can occupy relative to a fundamental Gaussian mode; only modes satisfying this criterion are transmitted, thereby fixing the subchannel count. The resulting approximations for each technique are[64]: $Q_{OAM}(S) \approx 2\text{floor}[S]+1$ for conventional OAM mode multiplexing, $Q_{LG}(S) \approx 0.5\text{floor}[S](\text{floor}[S]+1)$ for Laguerre-gaussian beam multiplexing, $Q_{HG}(S) \approx 0.5\text{floor}[S](\text{floor}[S]+1)$ for Hermite-Gaussian beam multiplexing, $Q_{MIMO}(S) \approx \text{round}[0.9S^2]$ for multi-input multi-output transmission, $Q_{IPVB}(S) \approx \text{floor}(189.16\sqrt{S-0.066})$ for Iso-propagation vortex beam multiplexing[30], respectively.

By employing the NDPVB basis—which inherently mitigates both diffraction-induced spreading and OAM-driven radial expansion (Fig. 5B)—a substantially expanded set of transmission modes can be accessed beyond those supported by conventional spatial multiplexing, thereby potentially boosting overall link capacity. Exploiting the OAM- and distance-invariant beam profile, the number of NDPVB subchannels is found, via the approach of ref. [64], to scale as $Q_{NDPVB}(S) \approx \text{floor}(569.21\sqrt{S}-3)$ (see Supplementary Note 8 for details). Figure 5C shows that, for realistic free-space links with finite apertures and $S < 30$, NDPVB multiplexing outperforms the existing schemes in available subchannels. For example, in our proof-of-principle setup ($S$ = 6.25, Supplementary Note 9), $Q_{NDPVB}$ reaches 1419—versus $Q_{OAM}$ = 13, $Q_{LG}$ = $Q_{HG}$ = 21, $Q_{MIMO}$ = 35, and $Q_{IPVB}$ = 471—an enhancement ranging from threefold to over a hundredfold. We further evaluated multi-vortex geometric (MVG) beam multiplexing[29], noting that for $S$ < 30, $Q_{MVG} \approx Q_{LG}$ and thus remains below $Q_{NDPVB}$. Although practical capacity also depends on mode-spacing choices and inter-channel crosstalk, a higher upper bound on $Q$ directly translates into a larger usable mode set and, typically, potentially increased capacity[22,23].

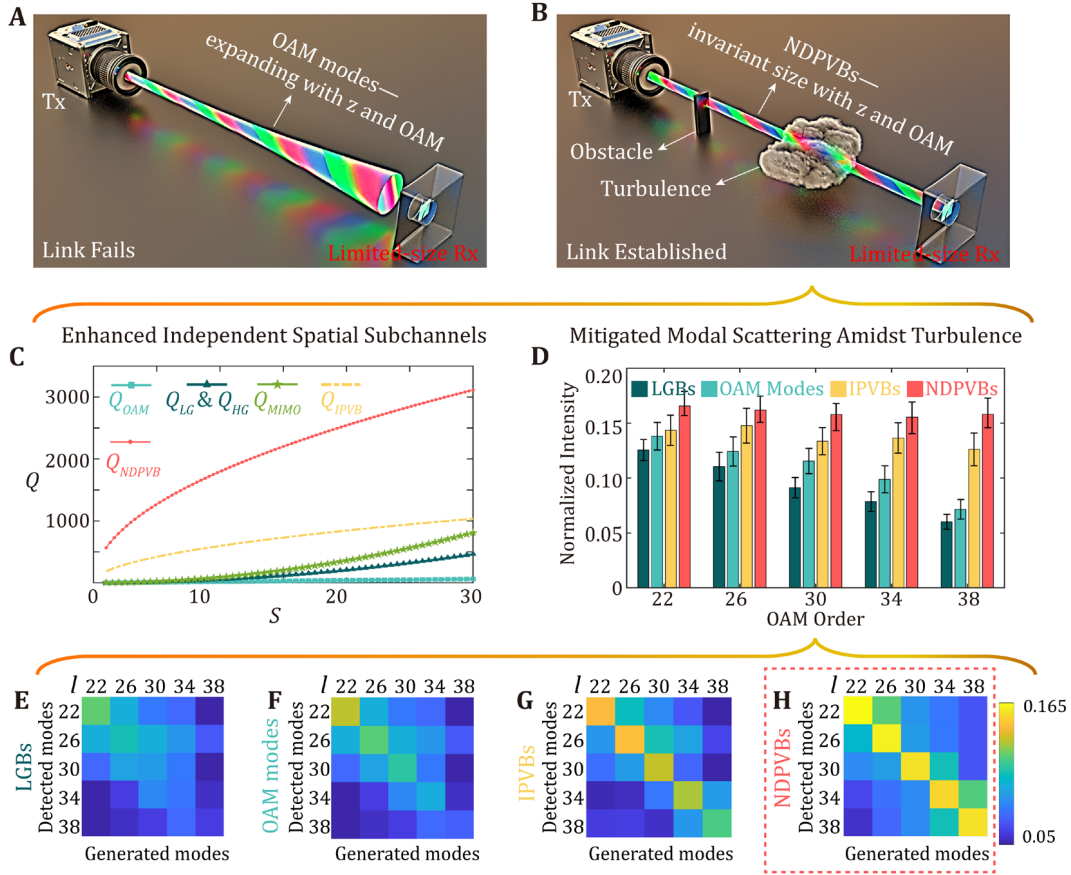

**Fig. 5. Enhanced Capacity and Mitigated Modal Scattering Amidst Atmospheric Turbulence. (A)** In conventional free-space communication, the limited-size receiver (Rx) obstructs OAM modes due to their expansion with propagation distance ($z$) and OAM order, resulting in **Link Fails**. **(B)** In contrast, NDPVBs maintain an on-demand, invariant size with respect to both z and OAM, enabling robust transmission and **Link Established** even in the presence of obstacles (self-healing ability) or turbulence. **(C)** Comparison of the number of independent spatial subchannels supported by various spatial multiplexing techniques, evaluated for system-quality factors ($S$) ranging from 1 to 30. **(D–H)** Assessment of free-space propagation under atmospheric turbulence for OAM beams, Laguerre-Gaussian beams (LGBs, nonzero–radial-index), iso-propagation vortex beams (IPVBs), and NDPVBs, respectively. The transmission distance is 1000 m with a turbulence strength of $C^2_n$ = 5×10$^{-15}$. Each beam has a consistent waist size while varying its OAM values ($l$ = 22, 26, 30, 34, 38). **(D)** Normalized intensity retained by each initiated mode at $z$ = 1000 m. **(E–H)** Crosstalk matrices for LGBs, OAM modes, IPVBs, and NDPVBs, providing a quantitative assessment of modal stability across the transmission pathway. Simulation details: propagation distance mesh of 40 m, analysis area of 0.5 m×0.5 m, and turbulence outer and inner scales of 300 m and 0.01 m, respectively. (See Supplementary Note 10 for further details.)

**Mitigate Modal Scattering in Turbulent Channels.** Atmospheric turbulence, together with diffraction-induced spreading and OAM-dependent radial expansion, limits both capacity and range in free-space links[1,22,23], and modal scattering grows with increasing turbulence strength or beam diameter increase[71]. As shown in Figs. 5D–5H, streamline-engineering NDPVBs, whose profiles are invariant to both OAM and propagation distance, exhibit lower and more uniform modal scattering than conventional OAM modes, Laguerre-gaussian beams and iso-propagation vortex beams under the same turbulent conditions. This turbulence resilience cements NDPVBs as potential carriers for high-capacity, turbulence-robust free-space multiplexing[66] (See Supplementary Note 10 for details). In addition to compensating for medium attenuation (Fig. 3), the controlled energy delivery to the NDPVBs' mainlobes could enable probing of atmospheric turbulence strength along the propagation path, thereby facilitating targeted mitigation of turbulent effects[67].

**Enable Robust NLOS Communication.** Free-space optical links—whose beam diameters are far smaller than those of radio waves—are prone to interruption by occluders, leading to severe signal loss or complete blackout. Conventional systems thus require an unobstructed line-of-sight (LOS) between transmitter and receiver (Figs. 5A–5B), which restricts their use in complex environments. By contrast, streamline-engineered NDPVBs exploit adjustable self-acceleration and self-similarity to steer around obstacles, extending robust transmission to non-line-of-sight (NLOS) scenarios under dynamic occlusions (Figs. 6A–6B).

Figures 6C-6N illustrates a high-dimensional NLOS proof-of-principle: a 128 × 128-pixel true-color "Mandrill" image (24-bit depth) was split into its red, green and blue channels (8 bits each) and mapped onto 24 NDPVB modes with indices $l$ = (-60, 60, -55, 55, … , -5, 5), using a mode spacing of 5 to suppress crosstalk. Two NLOS channel-engineering cases were tested—self-accelerating trajectory (Case 1) and self-similar radius (Case 2)—to navigate a dynamic occluder (red parts in Figs. 6E, 6G-6J). After demultiplexing (Supplementary Note 9 and Supplementary Movie 4), the image was faithfully reconstructed with bit-error rates of $2.19×10^{-4}$ (Case 1) and $8.78×10^{-5}$ (Case 2), both well below the $3.8×10^{-3}$ forward-error-correction limit (Fig. 6F). In contrast, conventional vortex beams and standard NDPVBs without streamline-engineering multimodal dynamics failed under the same conditions (Figs. 6K–6N). The setup—when using a digital mirror device at 11 kHz—achieved a peak data rate of 24×11k = 0.264 Mbit/s. Although our demonstration employed a single occluder due to current modulation limitations of commercial spatial light modulators, future high-performance platforms (e. g., metasurface) should enable even more complex NLOS links. Overall, streamline-engineered NDPVBs with multimodal propagation dynamics marks them as promising carriers for next-generation free-space optical networks[1,22,23] that combine ultrahigh capacity, turbulence resilience and robust NLOS transmission.

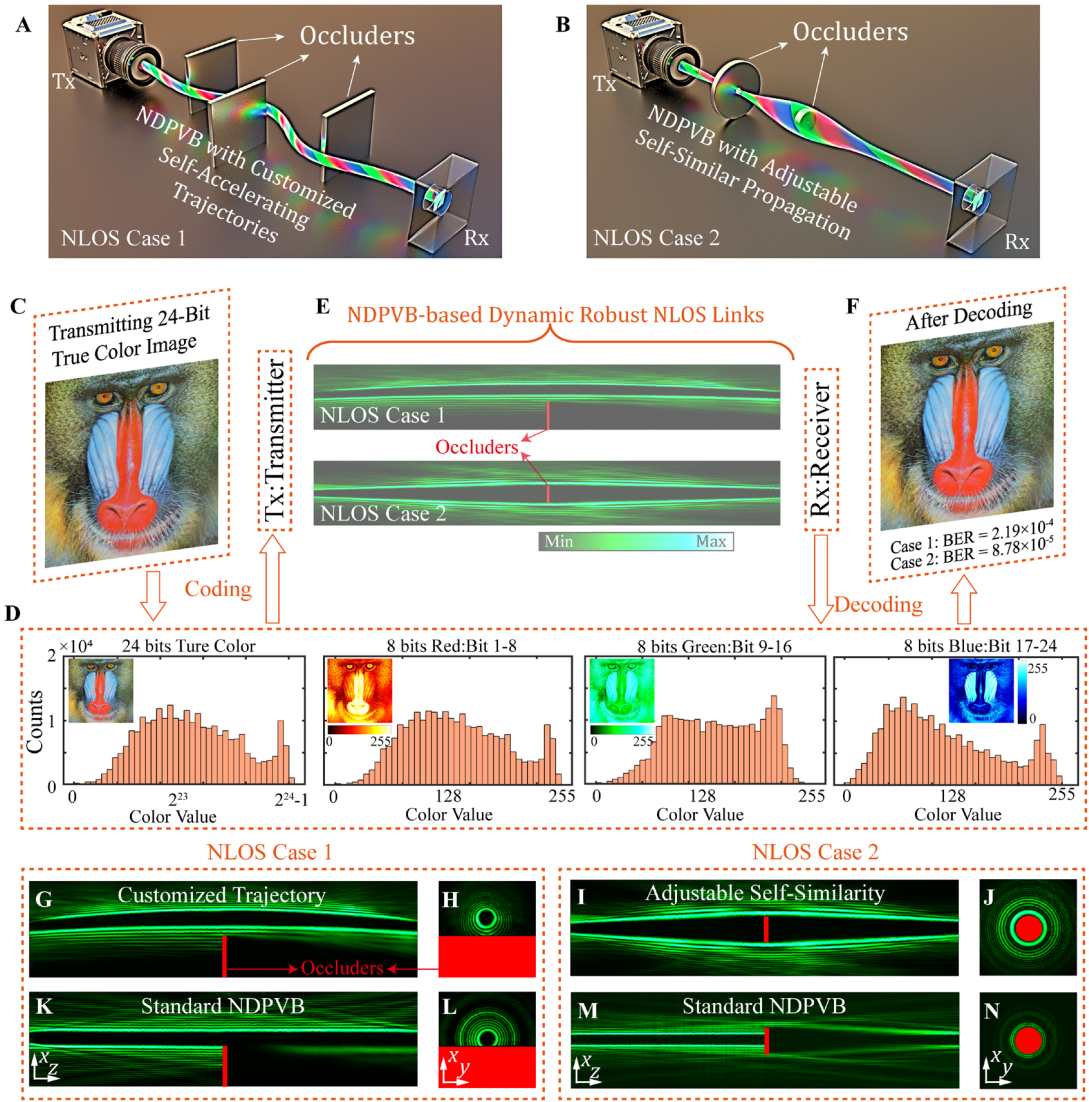

**Fig. 6. NDPVB-multiplexing NLOS image transmission with ultra-high color fidelity. A–B.** Schematics of non-line-of-sight channels: when conventional LOS links fail due to multiple occluders fully blocking the receiver, NDPVBs with (**A**) customized self-accelerating trajectories or (**B**) adjustable self-similar expansion navigate around obstacles to restore connectivity. Tx, transmitter; Rx, receiver. **C.** Original true-color "Mandrill" image (128 × 128 px, 24-bit depth, $2^{24}$ possible colors). **D.** Color histograms for the full image and its individual red, green and blue 8-bit channels, each mapped to bit-positions 1–24. **E.** Bit-stream mapping: 24 bits per pixel are encoded onto NDPVB modes with indices $l$ = (-60, 60, -55, 55, ... , -5, 5) (mode spacing = 5) and carried via (Case 1) self-accelerating or (Case 2) self-similar beams to bypass occlusions (red lines). **F.** Reconstructed "Mandrill" after demultiplexing, showing negligible distortion. Bit-error rates: $2.19 \times 10^{-4}$ (Case 1) and $8.78 \times 10^{-5}$ (Case 2), both well under the $3.8 \times 10^{-3}$ forward-error-correction threshold. **G–J.** Case 1 experimental intensity patterns at the occluder plane: (**G**) streamline-engineered beam self-bends to surmount a wall-shaped mask, (**K**) standard NDPVB is blocked. Corresponding x-y intensity maps are shown in (**H**) and (**L**), respectively. **K–N.** Case 2 experimental intensity patterns: (**I**) beam with expanded self-similar radius clears a plate-shaped mask, (**M**) standard NDPVB fails. Measured x-y intensity distributions appear in (**J**) and (**N**). See Supplementary Movie 5 for details of (**G–N**). The sharp-edged obstacles is produced as masks via the process of photoetching chrome patterns on a glass

substrate (Supplementary Note 11).

## Discussion

Building on the previous analysis, we have established a hydrodynamic framework for structured light by modeling the optical field as an energy "fluid" and tracking its three-dimensional evolution through energy streamlines. These streamlines provide an intuitive, three-dimensional roadmap of photon trajectories, capturing the dynamics of diffraction, OAM-induced expansion, self-healing, self-acceleration, and self-similarity within a unified framework. In contrast to classical structured modes, which follow fixed propagation laws and often require specialized techniques[14–17] to address individual propagation behaviors in isolation—insufficient for meeting the growing demands of real-world applications—the streamline-engineering approach offers a more flexible and unified strategy for controlling the multimodal propagation dynamics of structured light. By shaping the underlying streamline geometries, this framework addresses both diffraction- and OAM-induced broadening in vortex dynamics, while enabling customizable multimodal propagation that incorporates features from primary structured light families such as Bessel, Airy, and Gaussian beams.

Optical tweezer experiments, analogous to particle-tracking velocimetry in fluid systems, have been used to probe these streamline geometries. In these experiments, microspheres were observed to accurately trace the designed streamlines, indicating potential for three-dimensional optomechanical control via engineered photonic trajectories. This opens the door for applications such as microfluidic flow steering, directed cell transport, and the assembly of microscale structures[57]. The free-space communication case study with streamline-engineered multimodal NDPVBs demonstrated several improvements, including an increase in independent channels, reduced turbulence-induced mode scattering, and robust non-line-of-sight transmission. These findings suggest NDPVBs as promising carriers for next-generation optical networks[1,22,23]. Collectively, this streamline-driven framework bridges fluid dynamics and light propagation, offering a versatile, experimentally verified toolkit for controlling multimodal free-space dynamics. It opens new opportunities for structured-light applications[68] in imaging[52], manipulation[51], communication[22,23], and fluid dynamics simulations through optical analogues[69]. Additionally, this framework provides a potential tool for OAM research across vortex and vortex-free structured light modulities[48].

## Method

### 1. Four-step Streamline-Engineering Approach for Multimodal Propagation Dynamics

The principle of the four-step streamline-engineering approach is outlined in Table 1:

1. **Streamline Configuration**: Prescribe the desired trajectories $\boldsymbol{R}(z)$ to encode specific propagation behaviors.
2. **Momentum-field Sculpting.** Use the hydrodynamic equations [Eq. (3)] together with the fluid-continuity condition (Method 3) to compute the required momentum distribution $\boldsymbol{p}(\boldsymbol{R})$.

3. **Angular-spectrum engineering.** Devise an angular-spectrum distribution $S(k_r, \phi, k_z)$ in momentum-space that produces the target momentum field when transformed into real space.
4. **Real-space beam construction:** Implement an optical Fourier transform (via a lens) to convert $S(k_r, \phi, k_z)$ into the physical field $\psi(\mathbf{R})$, realizing deterministic control over its multimodal dynamics.

The four rows of Table 1 illustrate the manipulation of different propagation behaviors: (1) non-diffracting perfect vortex beams[70] with OAM- and z-invariant radius $r_l$, (2) adjustable self-similarity with dynamic radius $r_l(z)$, (3) customized self-acceleration with arbitrary trajectory $(x_s(z), y_s(z))$, and (4) tunable energy delivery with adjusted energy density $I(z)$. Detailed derivations for Table 1 can be found in Supplementary Note 1, and the validations of sculpted momentum fields can be found in Supplementary Note 2 with Supplementary Movie 6.

| Energy Streamlines $\mathbf{R}(z)$ Configuring | Momentum Fields $\mathbf{p}(r, \varphi, z)$ Sculpting | Momentum-Space Angular-Spectrum $S(k_r, \phi, k_z)$ Engineering | Real-space Multimodal Beam $\psi(\mathbf{R})$ Constructing |
|---|---|---|---|
| $\{r_l, \varphi_0 + l/(r_l^2 k_{zl})z, z\}$ | $\dfrac{l}{r_l}\hat{\boldsymbol{\varphi}} + k_{zl}\hat{\mathbf{z}}$ | $A_l e^{il\phi} \mathcal{F}_z\{e^{ik_{zl}z}\}$ | Non-diffracting Perfect Vortex Beams[64] with OAM- and z-invariant Radius $r_l$ |
| $\{r_l(z), \varphi_0 + l/[r_l^2(z)k_{zl}(z)]z, z\}$ | $kr_l'(z)\hat{\mathbf{r}} + \dfrac{l}{r_l(z)}\hat{\boldsymbol{\varphi}} + k_{zl}(z)\hat{\mathbf{z}}$ | $A_l e^{il\phi} \mathcal{F}_z\{\exp(i\int_{-\infty}^{z} k_{zl}(z)dz)\}$ | Adjustable Self-similarity with Dynamic Radius $r_l(z)$ |
| $\{x_s(z) + r_l\cos\varphi(z), y_s(z) + r_l\sin\varphi(z), z\}$ | $kx_s'(z)\hat{\mathbf{x}} + ky_s'(z)\hat{\mathbf{y}} + \dfrac{l}{r_l}\hat{\boldsymbol{\varphi}} + k_{zl}\hat{\mathbf{z}}$ | $A_l e^{il\phi} \mathcal{F}_z\{e^{ik_x x_s(z) + ik_y y_s(z)} e^{ik_{zl}z}\}$ | Customized Self-acceleration with Arbitrary Trajectory $(x_s(z), y_s(z))$ |
| Optical "Fluid-Continuity Equation" (Method 3) | $-CI'(z)\hat{\mathbf{r}} + I(z)\dfrac{l}{r_l}\hat{\boldsymbol{\varphi}} + I(z)k_{zl}\hat{\mathbf{z}}$ | $A_l e^{il\phi} \mathcal{F}_z\{\sqrt{I(z)} e^{ik_{zl}z}\}$ | Tunable Energy Delivery with Adjusted Energy Profiles $I(z)$ |

**Table 1. Four-step Streamline-Engineering Approach for Multimodal Propagation Dynamics.** $\hat{\mathbf{x}}, \hat{\mathbf{y}}, \hat{\mathbf{r}}, \hat{\boldsymbol{\varphi}}$ are the unit vectors in Cartesian and cylindrical coordinates, respectively; the single prime symbol denoting the first-order derivative with respect to z, e. g., $r_l'(z)$, $r_l'(z)$, $x_s'(z)$, $y_s'(z)$ and $I'(z)$. The constant $C$ in the fourth row is a positive constant, as will be discussed in Method 3. The subscript "z" of "$\mathcal{F}$" refers to the Fourier transform dimensions; $A_l = \sqrt[3]{3.2823|l| + 4.0849}$ represents the normalized amplitude coefficient; $k_{zl} \approx \sqrt{k^2 - ((|l|+2)/r_l)^2}$ and $k_{zl}(z) \approx \sqrt{k^2 - ((|l|+2)/r_l(z))^2}$ denotes the customized longitudinal wavevector component; (Supplementary Note 1).

## 2. Experimental Generation and Tomography for Multimodal NDPVB.

**Generation.** A reflective phase-only spatial light modulator imprinted with computer-generated hologram patterns transforms a collimated laser light wave into the complex field $S(k_r, \phi, k_z)$, with help of spatial filtering via a 4-F system consisting of lenses L1 and L2, and an iris as well (Fig. 7). The resulting field is responsible for multimodal NDPVB in the focal volume of lens L3. It should be noted that in the real-space coordinate system, the radial wavenumber $k_r$ must be converted into the radial coordinate in the incident plane (i.e. the front focal plane) according to the relation $k_r = kr/f$.

**Tomography**. A delay line, consisting of a right-angle and a hollow-roof prism mirrors and a translation stage, enables the different cross-sections of multimodal NDPVBs to be imaged on a CMOS camera after a relay 4-F system consisting of two lenses (L4 and L5). The combination of the delay line and the relay system (Fig. 7) enables us to record intensity cross sections at different z-axial locations in the focal volume of lens L3. (The actual experimental apparatus provided in Supplementary Note 3)

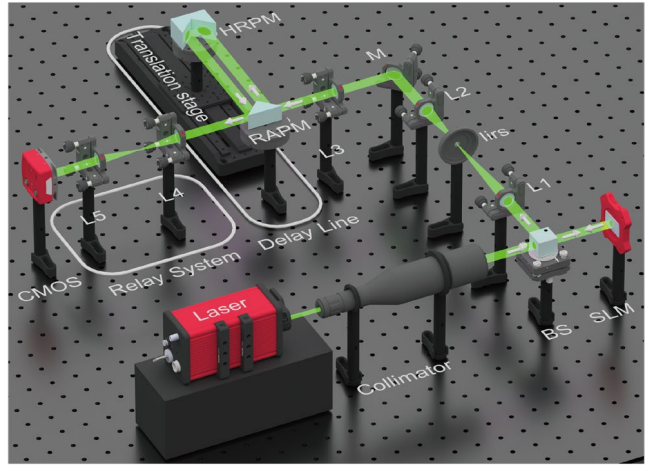

**Fig. 7. Experimental setup for Generation and Tomography of Multimodal NDPVB.** BS, beam splitter; SLM, phase-only spatial light modulator; L1-L5, lens; M, mirror; RAPM, right-angle prism mirror; HRPM, hollow roof prism mirror (LBTEK HRM25-AG).

3. **Streamline Interaction Manipulation by Optical "Fluid-Continuity Equation".**

The propagation of optical energy in a beam can be described by optical "Fluid-Continuity Equation"[41] (namely, Transport of Intensity Equation[71]), read as:

$$\frac{\partial W(r,\varphi,z)}{\partial z} = -c(\nabla_\perp \boldsymbol{p}_\perp(r,\varphi,z)), \quad (7)$$

where $W(r, \varphi, z)$ is the energy density, $\boldsymbol{p}_\perp(r, \varphi, z)$ the transverse energy-flow (momentum) density, $\nabla_\perp$ denotes the transverse divergence operator, and $c$ is the velocity of light. This relationship indicates that the transverse energy-flow (momentum) density $\boldsymbol{p}_\perp(r,\varphi,z)$, influences the propagation of optical energy density $W(r, \varphi, z)$. If one treats $W$ as a fluid density field and $\boldsymbol{p}_\perp$ as the corresponding transverse velocity field, Eq. (7) becomes exactly the continuity equation of incompressible-fluid dynamics, enforcing conservation of matter.

Non-diffracting perfect-vortex beams (NDPVBs) feature an invariant mainlobe, regardless of topological charge $\ell$ or propagation distance $z$. We therefore decompose these beams into two regions: mainlobe (region of interest): the high-intensity region that interacts directly with matter or detectors; sidelobes: a concentric series of rings that sustain and regulate the mainlobe's propagation. During free-space propagation, energy streamlines of standard NDPVB in these two regions $\{r, \varphi_0 + l/(r^2 k_{zl})z, z\}$, circulate on distinct cylindrical surfaces without exchanging energy. However, by invoking Eq. (6), we can actively couple the sidelobes to the mainlobe as an on-demand energy reservoir.

Considering the ring area of the mainlobe at $r= r_1$, the derivative $\partial W(r_1, \varphi, z)/\partial z$ is directly proportional to $\partial I(z)/\partial z = I'(z)$. The divergence of transverse energy flow or momentum density, $\nabla_\perp \mathbf{p}_\perp(r, \varphi, z)$, can be equated to the flux of transverse energy flow across the ring. Owing to the beam's perfect axisymmetry, azimuthal flows circulate within the ring and do not contribute to net flux; only the radial component matters. Hence, the radial energy flow exhibits a direct correlation with $-I'(z_0)$, donated as $-CI'(z_0)$ with the integrated positive constant $C$ (Raw 4, Table 1). When $I'(z)>0$, the negative sign in radial energy flow indicates an inward radial flow: sidelobe streamlines converge into the mainlobe, replenishing and amplifying its energy (see Fig. 3C). When $I'(z) < 0$, the flow reverses, allowing controlled depletion of the mainlobe back into the sidelobes.

By programming the spatial light modulator to impose a desired axial energy gradient $I'(z)$, one gains *continuous, bidirectional* control over the sidelobe-to-mainlobe energy transfer rate. This tunability enables:

1. Enhanced local intensity—by funneling nearly all sidelobe energy inward, one can boost detector signal-to-noise and overall energy efficiency (Fig. 3C, a three-fold improvement is detailed in Supplementary Note 6).
2. Attenuation compensation—by matching the sidelobe streamlines inflow rate to medium-induced decay, the mainlobe's intensity remains constant during propagating (Figs. 3G–J).

In this way, the sidelobes serve not only as passive scaffolding for self-healing (Fig. 3A) but also as an *actively addressable energy reservoir* for precision-tailored structured-light applications[52] (Figs. 3G–J).

## Acknowledgements
**Funding:**

National Key Research and Development Program of China (2023YFA1406903),

National Natural Science Foundation of China (12374307, 12234009, 12427808)

**Competing interests:**

The authors declare no competing interests.

**Data and materials availability:**

All data that support the findings of this study are available in the main text and the supplementary materials, or available from the corresponding author on reasonable request.